\documentclass{elsart3}

\usepackage{natbib}

 \usepackage{graphicx}

\usepackage{amssymb}

\begin{document}

\begin{frontmatter}



\title{Star Formation in QSO Host Galaxies\thanksref{hst}}
\thanks[hst]{Based in part on observations made with the NASA/ESA Hubble Space 
Telescope, obtained at the Space Telescope Science Institute, which is 
operated by the Association of Universities for Research in Astronomy, Inc., 
under NASA contract NAS 5-26555. These observations are associated with 
programs \# GO-10421 and \# SNAP-10588}

\author[label1]{Gabriela Canalizo}
\author[label2]{Alan Stockton}
\author[label3]{Michael S. Brotherton}
\author[label4]{Mark Lacy}

\address[label1]{Department of Physics and Institute of Geophysics and Planetary Physics, University of California, Riverside}
\address[label2]{Institute for Astronomy, University of Hawaii}
\address[label3]{Department of Physics and Astronomy, University of Wyoming}
\address[label4]{Spitzer Science Center, Caltech}



\begin{abstract}
Many of the conditions that are necessary for starbursts appear to be 
important in the triggering of QSOs.  However, it is still debatable whether 
starbursts are ubiquitously present in galaxies harboring QSOs.
In this paper we review our current knowledge from observations
of the role of starbursts in different types of QSOs.  
Post-starburst stellar populations are potentially present in 
the majority
of QSO hosts.   QSOs with far-infrared colors similar to those
of ultraluminous infrared galaxies invariably reside in merging galaxies 
that have
interaction-induced starbursts of a few hundred Myr or less.  Similar,
but dramatically more luminous post-starburst populations are found in
the recently discovered class of QSOs known as post-starburst QSOs, or Q+A's.
Both of these classes, however, comprise only a small fraction ($10-15$\%) of 
the total QSO population.   The so-called ``red'' QSOs generally suffer
from strong extinction at optical wavelengths, making them ideal candidates for
the study of hosts.   Their stellar populations typically show a post-starburst
component as well, though with a larger range of ages.
Finally, optical ``classical'' QSO hosts show traces of 
major star formation episodes (typically involving $>$10\% of the mass of 
the stellar
component) in the more distant past (1-2 Gyr).  These
starbursts appear to be linked to past merger events.  It remains to be 
determined
whether these mergers were also responsible for triggering the QSO activity
that we observe today.

\end{abstract}


\end{frontmatter}

\section{Introduction}
\label{Intro}
The relationship between mergers, starbursts, and active galactic nuclei (AGN)
has long since been the subject of vigourous research.   It is clear that at 
least some of the ingredients necessary to trigger starbursts are also 
necessary to trigger AGN: both phenomena require fuel, and both require that
this fuel be somehow displaced from one point to another, whether to replenish
the material in the accretion disk, or to compress it resulting in enhanced
star formation.   There is also evidence that mergers are important to both 
phenomena, although the precise role that they play is still debatable. 
The high incidence of mergers in starburst galaxies \citep[][and references 
therein]{san96} might indicate that mergers are indeed necessary to trigger 
massive
starbursts, but they are not sufficient, since mergers often result in only
moderately enhanced star formation \citep[e.g.,][]{ber03}.  Mergers have,
on the other hand, only been shown to be required to trigger nuclear 
activity in far-infrared (FIR) loud QSOs \citep{can01}, and are certainly not
sufficient, as can be seen from the relatively low incidence of AGN in any 
catalog of interacting galaxies.

What are the additional ingredients necessary to trigger starbursts and AGN?
Are these ingredients common to both phenomena?  We already know that 
starbursts can exist in the absence of AGN, as has been found in the case of 
many ultraluminous infrared galaxies \citep[at least at the lower luminosity 
end; e.g.,][]{san96}.   In these proceedings, we address the question of 
whether the
converse is true, i.e., whether it is possible to trigger a high luminosity
AGN (specifically a QSO) without triggering also a starburst.   In the 
following sections, we review our 
current knowledge of star formation in different kinds of QSOs.  For the
purposes of this discussion, have 
grouped QSOs in four classes according to some of their observational 
characteristics rather than their intrinsic properties.

\section{Starbursts in QSO Hosts}
\subsection{Color-Selected FIR QSOs}
\label{firqsos}

One hypothesis that clearly ties in merger-induced star formation with QSO 
activity is that of \citet{san88}, who suggest that ultraluminous infrared
galaxies (ULIRGs) play a dominant role in the formation of all QSOs.  
According to this hypothesis, ULIRGs are the result of strong interactions 
or mergers which funnel gaseous material into the central regions of galaxies,
thus fueling intense star formation and the QSO activity.  
ULIRGs are then dust-enshrouded QSOs which, after blowing away
the dust, become classical QSOs.

\begin{figure}
{\centerline{\includegraphics[width=3.1in]{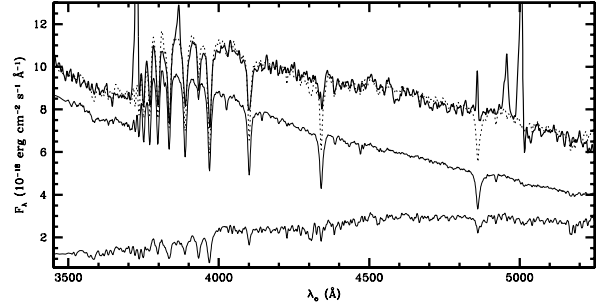}}}
\caption{Spectrum of the FIR QSO 3C\,48 host galaxy in a region $\sim$2'' E of
the quasar (heavy solid line).  This spectrum has been fitted with a 
Bruzual-Charlot model (dotted
line) comprising a starburst component of weighted average age 5 Myr
(upper light solid line)
and an older component with exponentially decreasing star formation
(e-folding time 5 Gyr; lower light solid line).
The lower Balmer absorption lines are contaminated with emission and
are not used in the fitting procedure; these and other emission lines are due, 
not to H\,II regions, but rather to extended emission photoionized by the 
quasar.
The exposure totaled 3600 s with LRIS on Keck II through a 1'' slit.
\citep[From][]{can00b}.
\label{3c48spec}}
\end{figure}

We have tested this hypothesis through a systematic imaging and spectroscopic
study of host galaxies of low-redshift QSOs
found in a region of the far-infrared (FIR) two-color diagram between the 
region where most QSOs lie and the region occupied by ULIRGs
\citep[][and references therein]{can01}.  These objects are presumably in 
some transition stage
between the ULIRG and ``normal'' QSO phases.  
Spectra were obtained of the host galaxies and/or strongly interacting
companions for all QSOs in the sample with the Keck Low-Resolution 
Imaging Spectrometer (LRIS).  We obtained ages for the 
starburst or post-starburst component in different regions of each host 
galaxy using the procedure illustrated by Fig.~\ref{3c48spec}.  Using
these data along with $HST$ and ground-based 
images, we constructed detailed star-formation 
and interaction histories for each of the objects in the sample.  

We found that every ``transition'' QSO is undergoing a strong 
tidal interaction, and most are major mergers where
at least one of the parent galaxies was a disk galaxy.
The spectra are 
characteristic of E+A galaxies, and are 
successfully modeled by an underlying old population (the 
stellar component present in the parent galaxies prior to interaction)
plus a superposed instantaneous burst population (presumably resulting
from the interaction).  
All of the hosts have very young starburst ages,
ranging from ongoing star formation to $\sim300$ Myr.
By modeling spectra from many discrete regions across the
hosts, we created velocity fields and age maps from the 
stellar populations.
By comparing the starburst ages of the central stellar components
with those of the more extended emission, we determined the relative
ages between stellar populations in various regions of the host galaxies.
These estimates, along with dynamical ages, place constraints on the
timescale for concentrating material in the nucleus.
The concentration of material is likely to have 
triggered the central strong starbursts and the QSO activity roughly 
simultaneously.  The age of the peak starburst is, therefore, 
representative of the age of the QSO activity.  
To summarize, our study showed that the QSO and ULIRG phenomena
are physically related in these transition objects, and firmly established
that at least some QSOs can be traced back to a merger and a starburst
phase.

Star formation in these QSOs is then, not only prominent, but clearly linked
to the triggering of the QSO activity.   However, these FIR QSOs
may not be telling the story of the QSO population as a whole.   By comparing
the distribution of the ratio $L_{IR}/L_{UV}$ for PG QSOs \citep{san89} and 
for ``transition'' QSOs, we estimate that the transition sample is
representative of $\sim7.5\%$ of the optically selected PG QSO sample.
While there are many uncertainties in this estimate (especially since the
ratio may be more indicative of reddening
than of intrinsic characteristics in the objects), it clearly shows that 
these objects
are not representative of the majority of (at least optical) QSOs, but 
rather form a relatively
small fraction of the population.

\subsection{Post-starburst QSOs}

The rare class of post-starburst quasars shows simultaneously an AGN and a
massive luminous starburst of a few $\times$100 Myr old.  Because of their
composite spectra displaying broad emission lines as well as the Balmer jump 
and strong Balmer absorption lines characteristic of type-A stars, these
QSOs are sometimes called ``Q+A'' objects.  A striking 
case is UN J1025$-$0040 \citep{bro99}.
The strong starburst component has an age of $\sim400$ Myr, and a
bolometric luminosity of $10^{11.6}$L$_{\odot}$, equal to that of the quasar.
A younger UN J1025$-$0040 (tens of Myr after the
starburst) would have a more luminous stellar population and would likely
be dust enshrouded, placing it in the ULIRG class.  UN J1025$-$0040 has a 
nearby companion galaxy also in a post-starburst phase \citep{can00}.
The companion appears to be interacting with the quasar host galaxy, and this 
interaction may have triggered both the starburst and the quasar activity in 
UN J1025$-$0040. $HST$ WFPC2 imaging \citep{bro02} shows that the 
starburst is nuclear, the host resembles a merger remnant, and a less massive 
young starburst is also present.

\begin{figure}[b]
{\centerline{\includegraphics[width=3.0in]{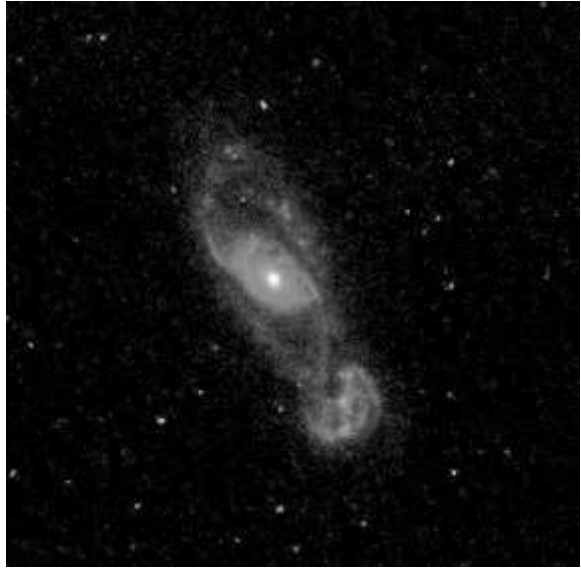}}}
\caption{$HST$ ACS snapshot of the post-starburst 
QSO SDSS\,231055$-$090107 and its interacting companion galaxy. \label{qaimg}}
\end{figure}

\begin{figure}
{\centerline{\includegraphics[width=3.3in]{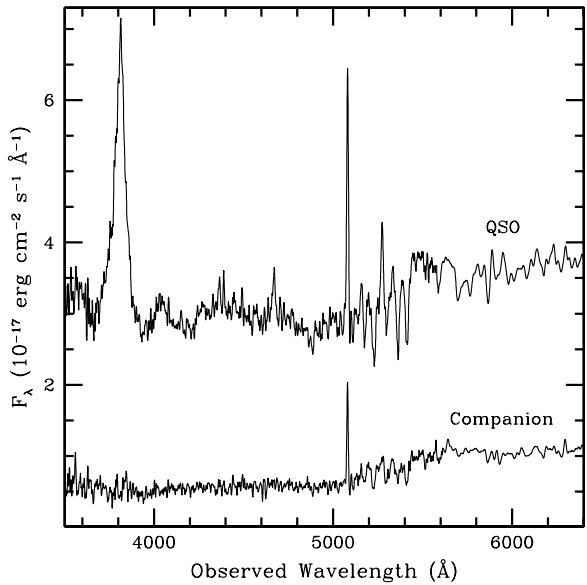}}}
\caption{Keck LRIS spectra of the post-starburst 
QSO SDSS\,231055$-$090107 an its companion, showing that 
both have undergone dramatic episodes of star formation. \label{qa}}
\end{figure}

The new large QSO surveys 2dF and SDSS are revealing post-starburst QSOs 
in significant numbers for the first time.  We have spectroscopically 
selected post-starburst quasars in the SDSS data release 2 (DR2) using an 
automated algorithm based on one that \citet{zab96} used to 
select post-starburst galaxies.
We have found $\sim$250 post-starburst 
quasars -- roughly 5\% of $z < 0.75$\ QSOs -- that show clear evidence 
for post-starburst stellar populations of significant luminosity/mass
(Brotherton et al, in preparation).  
SDSS images show that about 40\% of 
our objects appear 
to be close interacting doubles or have companions within 
10\hbox{$^{\prime\prime}$}.  
Even most single sources show some extended fuzz from 
the host galaxy, and often features like tidal tails indicating merger 
activity.   An $HST$ Advanced Camera for Surveys (ACS) snapshot survey of 
these objects is currently 
underway and is already revealing a much higher rate of tidal interaction
than can be inferred from the SDSS images.  In Figs.~\ref{qaimg} and 
\ref{qa} we show, respectively, the 
image and spectra of one of the SDSS QSOs with its companion, which appears to 
have a similar, though much
less luminous, post-starburst population to that of the host.

Post-starburst QSOs represent the most dramatic cases of star formation in QSO
host galaxies.  However, they are only a small fraction of the total QSO
population, and their relationship to other QSOs as well as to ULIRGs and
E+A galaxies is currently unknown.   We are conducting 
a comprehensive study including $HST$ ACS (Fig.~\ref{qaimg}), Keck 
LRIS (Fig.~\ref{qa}), and Spitzer IRS observations
that will help us determine the nature of this relationship.

\subsection{Red QSOs}
Another class of QSOs that is being uncovered with the advent of large area 
surveys at longer wavelengths, such as 2MASS and FIRST, is that of the 
so-called ``red'' QSOs.   In general, these objects have some of the same
characteristics as blue QSOs, such as (at least some of the) strong broad 
emission lines and high
bolometric luminosities, but with much redder continua.  At present, there
is not a clear definition for red QSOs, so that the different objects that
are catalogued as red QSOs do not form a a homogeneous class.
For example, in the sample of red QSOs drawn from the Parkes Half-Jansky 
Flat-radio-spectrum Sample \citep[PHFS;][]{fra01}, the reddening is most 
likely due to red synchrotron emission.  
The majority of red QSOs discovered to date, however, appear to be 
reddened by dust, and thus they are considered the dust-obscured equivalent
of the blue QSO population \citep[e.g.,][]{cut02,mar03,hal02,gli04,whi03}.
Even then, there are cases where the dust is intervening rather than
intrinsic to the QSO \citep[e.g.,][]{gre02}.
Here we focus on those objects where the dust is presumably near the nucleus.

The nature of red QSOs remains uncertain. There are two popular theories to 
explain why some quasars appear reddened by dust. The first uses an analogy 
with the Seyfert galaxy and radio galaxy/quasar orientation-based unified 
schemes in which whether the observer sees a type-1 ( Seyfert-1, normal 
quasar) object or a type-2 (Seyfert-2, quasar-2) object depends on the angle 
at which the observer views the AGN. In this model, dust-reddened quasars are 
predicted to be objects viewed close to the transition angle between type-1 
and type-2 which just scrapes the surface of a torus of gas and dust 
surrounding the AGN. The radio-loud red quasar 3C68.1 is thought to be at an 
angle to the line of sight between those of normal radio-loud lobe-dominated 
quasars and their fully obscured counterparts, radio galaxies 
\citep{bro98}.
The second theory is that the dust which reddens the QSO is produced in 
young objects as a result of the starburst which follows the merger of two 
galaxies and which also triggers the QSO, once again as in the scenario of 
\citet{san88}. Thus a red QSO phase might be expected at some point early 
in the life of most quasars.

\begin{figure}[hb]
{\centerline{\includegraphics[width=3.1in]{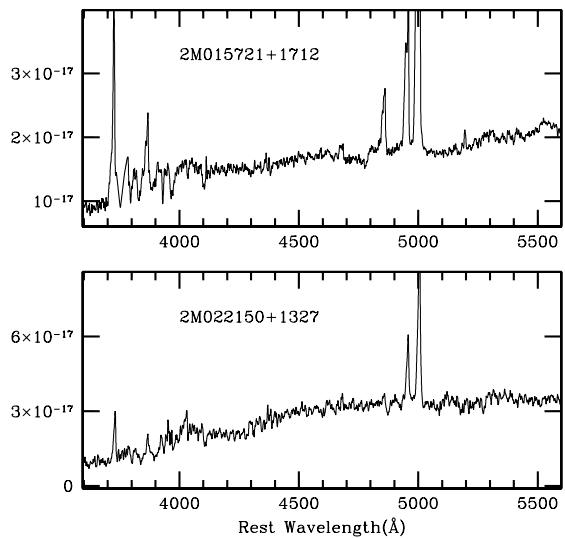}}}
\caption{Keck ESI spectra of red 2MASS QSOs.   Although these 
objects show broad H$\alpha$ emission characteristic of QSOs, the spectra of 
the host galaxies suffer little contamination from the QSO at shorter 
wavelengths. The spectra shown here have not been corrected for scattered QSO 
light to illustrate this point.   The spectrum of 2M015721+1712 shows a 
post-starburst spectrum characteristic of post-starburst QSOs while that of
2M022150+1327 shows a somewhat older population.\label{redspec}}
\end{figure}

In either case, the heavy extinction of the QSO nuclei at optical wavelengths 
makes red QSOs
excellent candidates to study stellar populations in their host galaxies,
yet this fact has remained largely unexploited.   We have obtained Keck ESI
spectra of 15 red QSOs from the \citet{mar03} 2MASS sample; spectra of two
of these objects are shown in Fig.~\ref{redspec}.  
The spectra, which clearly 
show stellar absorption lines, have not been corrected for scattered QSO 
light, demonstrating that they suffer little contamination 
from the nucleus at wavelengths shorter than H$\alpha$.  

Roughly half of 
the sample show strong post-starburst populations similar to those of 
post-starbursts QSOs.  Of the remaining objects, most, if not all, show 
evidence for relatively recent star formation.  Thus it appears that 
starbursts also play an important role in these objects.

How common are red QSOs?  \citet{gli04} estimate that heavily dust-reddened
QSOs comprise 20\% of the total QSO population for $K < \sim 15.5$.
\citet{lac02} and \citet{whi03}, among others, 
suggest that so far we have only detected the 
``tip of the red quasar iceberg'', which could comprise up to several times
the blue QSO population.   As more surveys are published and selection methods
are refined, estimates for the fraction of red QSOs are likely to become
more accurate.  This will have important consequences in areas ranging 
from galaxy evolution to cosmology and, for the current discussion, it will
allow us to have better statistics on the number of QSO hosts that show
young post-starburst populations.

\subsection{Classical QSOs}

We use the term ``classical'' here to describe any luminous optical QSO
not covered by the first three classes of QSOs discussed above.  
Recent imaging studies have shown that the majority of low-redshift 
classical QSOs, whether radio-loud or radio-quiet, reside in 
the centers of galaxies that have relaxed light distributions like 
those of elliptical galaxies \citep[e.g.,][]{dis95,bah97,dun03,flo04}. 

Unlike the case of red QSOs, the study of stellar populations in classical
QSO hosts is hampered by the difficulty of observing absorption line
spectra in galaxies that tend to be overpowered by their bright nuclei.
For this reason, there have been few spectroscopic studies that deal with
star formation in these objects.

\citet[][and references therein]{dun03} carried out imaging and 
spectroscopy of a statistically-matched sample of radio galaxies, quasars, and 
radio-quiet QSOs at $z\sim0.2$.
They find that essentially all of the host galaxies having nuclei in
the quasar luminosity range are bulges that have properties 
``indistinguishable from those of quiescent, evolved, low-redshift 
ellipticals of comparable mass.''   In particular, they suggest that 
these galaxies have truly old stellar populations with no episodes of 
massive star formation in the recent past.

Several other surveys of AGN host galaxies \citep[e.g.,][]{jan04,san04,kau03} 
also indicate that galaxies hosting the most luminous AGN are most often 
bulge-dominated.  However, all of these surveys find that the host colors 
are significantly bluer than those of inactive elliptical galaxies and they
are consistent with the presence of intermediate age starbursts.   Based on
their position in the $D_{n}$(4000)/H$\delta_A$ plane, \citet{kau03} suggest
that these AGN hosts have had significant bursts of star formation in the
past $1-2$~Gyr.  

\begin{figure}
{\centerline{\includegraphics[width=3.4in]{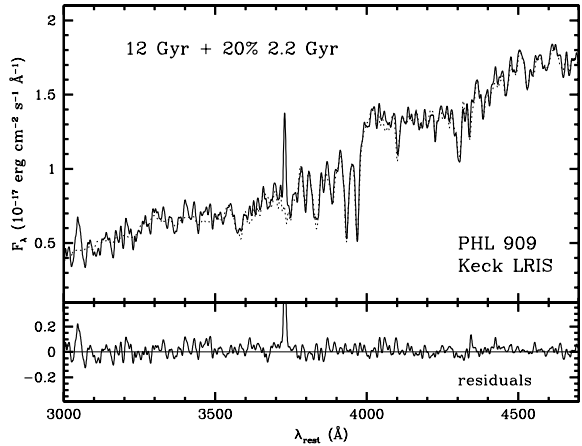}}}
\caption{Keck LRIS spectrum of the host galaxy of the classical QSO 
PHL 909 in rest frame 
(heavy line) and a $\chi^2$ fit of a two component model to the data 
(dotted line).  The model is the sum of a 12 Gyr old population
and a 2.2 Gyr instantaneous starburst model which 
comprises 40\% of the mass of the total population.  The bottom panel shows 
residuals of the fit.\label{oldsb}}
\end{figure}

The objects studied in these surveys are on average 
$\sim1$ magnitude fainter than those in the Dunlop et al.\ sample.  However,
it is possible that the same signs of intermediate age starbursts may be 
present in the hosts of higher luminosity objects, but that they are
simply more elusive due to the technical difficulties mentioned above.  To 
investigate
this possibility, we have obtained and modeled 
very deep ($\sim2$ hr exposures) Keck LRIS 
spectra of 14 $z\sim0.2$ classical QSO host galaxies
in the Dunlop et al.\ sample that are dominated by spheroids.
The surprising results are that we
have found a starburst component in all but one of the host galaxies, and in
the majority of objects we find traces 
of major starburst episodes involving $>10\%$ of the mass of the stellar 
component, with ages ranging from 0.6 Gyr to 2.2 Gyr (see Fig.~\ref{oldsb}).
While there are uncertainties in the 
precise starburst age estimates due to metallicity, initial mass function, 
and the models themselves, the high signal-to-noise ratio of the spectra 
clearly show, in addition to an underlying very old component, a younger 
stellar component that is significantly older than a few $\times10^8$ yr. 
This component cannot be mimicked either by a small amount of constant
star formation or by one less dramatic and substantially younger starburst 
phase.   In any case, our results show conclusively that the host galaxies of
classical QSOs are $not$ made up almost purely of ancient stellar populations.

\begin{figure}
{\centerline{\includegraphics[width=3.0in]{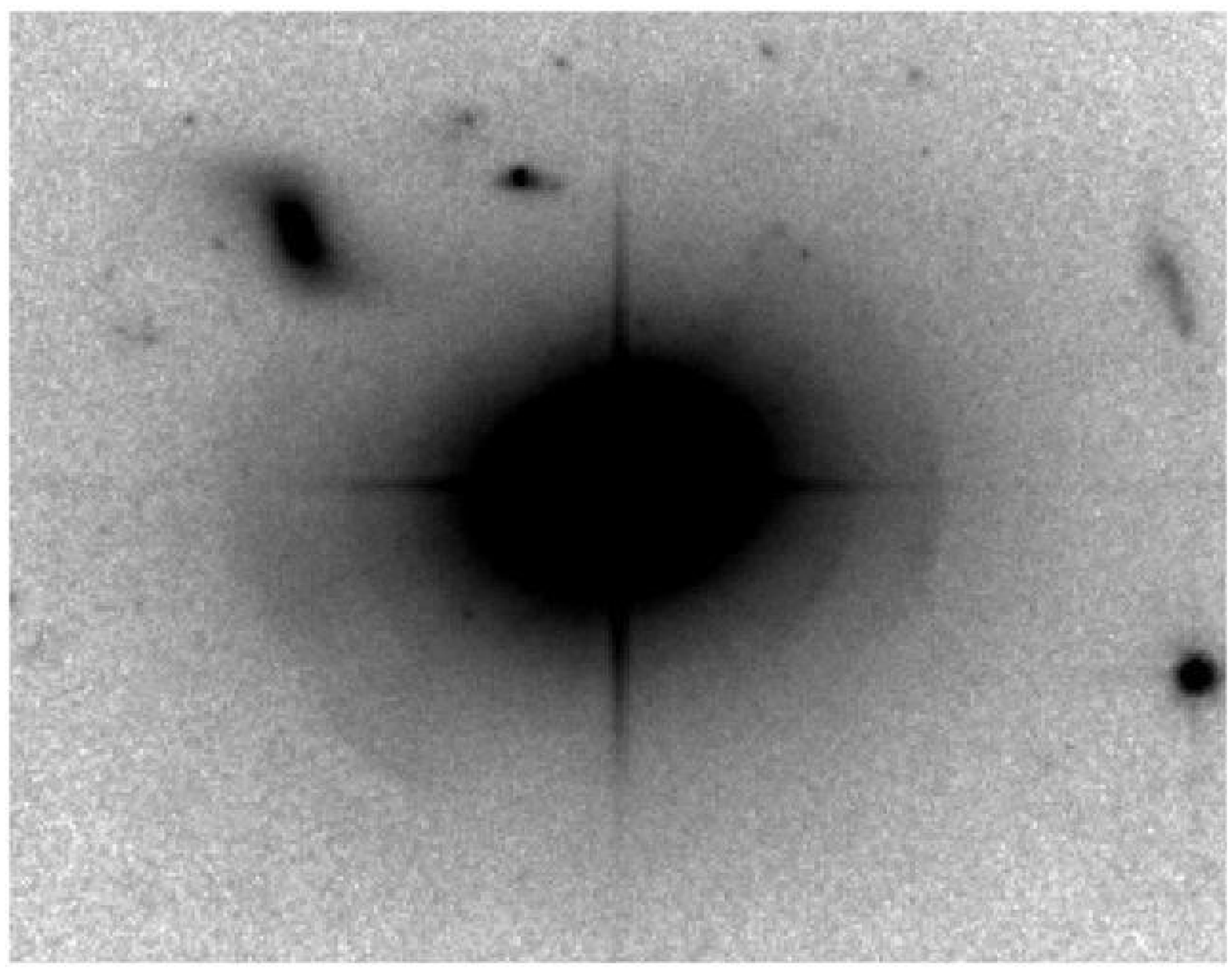}}}
{\centerline{\includegraphics[width=3.0in]{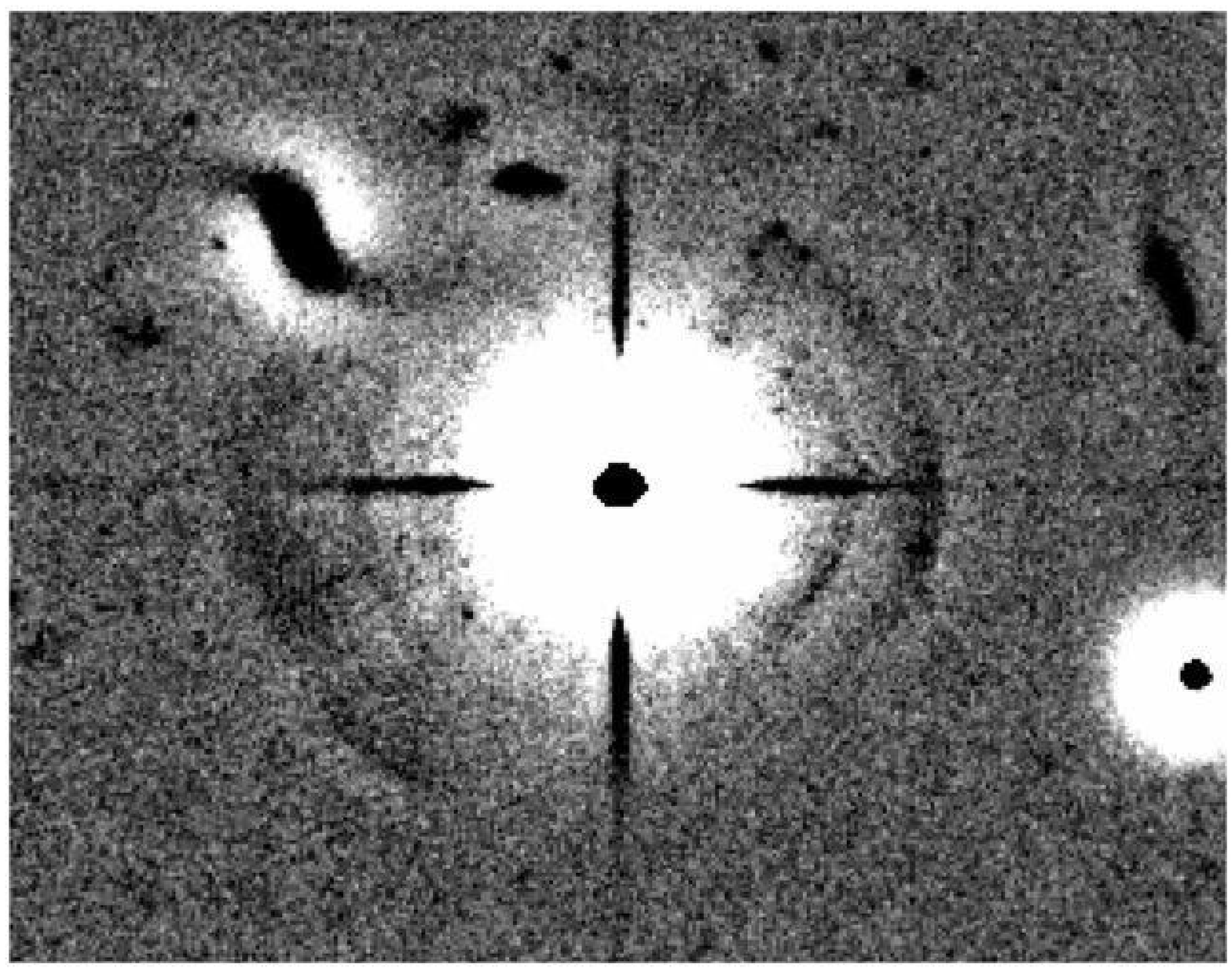}}}
\caption{$HST$ ACS image of the $z\sim0.2$ classical QSO MC2 1635+119 
showing spectacular
shell structure.  The second panel shows unsharp-like masking to highlight the 
sharp features that make up the shells.\label{shells}}
\end{figure}

Are these major starbursts episodes connected to a merger event?  
Elliptical hosts formed through mergers would be expected to
show fine structure indicative of past tidal interactions.  
This tell-tale structure, which includes shells, ripples, 
and boxy isophotes, is commonly seen in nearby ellipticals 
and indicates that these early type galaxies were either 
formed or structurally modified by mergers in the not so distant
past.  Fine structure
tends to persist even after the more overt signs of interaction have faded,
and it is thought to be sensitive to the dynamical age of the merger
\citep[e.g.,][]{sch90,sans00}.
These studies indicate that fine structure can in general be detected
even a few Gyr after the last major merger event.  
We are obtaining very deep $HST$ images of five of the host galaxies for which we have Keck spectra, using ACS with the F606W broad-V filter. 
The subsample was chosen to include those objects for which we have
the most robust age determinations and that do not show emission
lines in the host galaxy to avoid confusion in the interpretation of the 
imaging when using a broad filter.  Three objects have been observed so far, 
and all three show signs of tidal interactions: one appears to be in a system
of three or four interacting galaxies, while the other two show traces of 
past major merger events.   Figure~\ref{shells} shows a dramatic example of 
shell structure in one of the QSO hosts.

The results from both our Keck spectroscopic and $HST$ imaging 
observations point toward the same conclusion: classical QSOs reside in host
galaxies that have had a major tidal interaction and starburst episode 
within the
last two or so billion years.   But the question remains:  Is there a
connection between this catastrophic event in the past and the current 
nuclear activity?   An oversimplistic interpretation assuming that the
Gyr old merger triggered the QSO activity at that time implies a QSO duty 
cycle of more than a billion years, a result in stark 
contrast to current theoretical  estimates 
\citep[e.g., $(3-13)\times10^{7}$ yr by][]{yu02}.   
The scenario may be more complicated, however.
QSO activity may require, for example, a two step process: first, collection of
gas by a major merger, and then, triggering of the nuclear activity by other
mechanisms such as a minor merger.

However, before we worry about these complex possibilities, a more fundamental 
question needs to be considered:  are the host galaxies of classical QSOs 
truly distinct from typical inactive elliptical galaxies?  In other words, if 
we were to study a sample of inactive ellipticals at $z\sim0.2$ 
in the same depth, might we find similar shell structure and traces to major 
starburst episodes?  A matching study of a control sample of elliptical 
galaxies in environments similar to those of QSOs and with similar effective 
radii is clearly necessary in order to answer these questions.
Such a study would allow us to interpret our results in the proper context
and determine once and for all whether classical QSOs reside in normal 
elliptical galaxies or not.

\section{Summary}

Although we are not yet able to answer the question of whether the
triggering of a QSO is necessarily accompanied by the triggering of
a starburst, we have strong hints that this may well be the case.
Star formation plays a prominent role in
three of the QSO classes we have discussed: FIR QSOs, Q+A's, and
red QSOs. Most of these objects, as well as a small fraction of classical 
QSO hosts,
show relatively young post-starburst populations of a few $\times$10$^8$ yr.  
However, is still unclear what fraction of all QSOs is
represented by these objects.   A conservative estimate would indicate that
they form at least one fourth of the total QSO population if we 
add 7.5\% for FIR QSOs, 5\% for Q+A's, and 20\% for red QSOs, and allow
for some overlap between these classifications.  However, the fraction could
be as high as 80\% or more if the population of red QSOs proves to be as
numerous as currently suspected \citep{whi03}.
The rest of the QSO hosts appear to have traces of massive starbursts 
with ages $(1-2) \times$10$^9$~yr.  Whether these starburst episodes are linked
to the event that triggered the nuclear activity remains to be established.



\bigskip

\noindent
{\bf Acknowledgments}

\bigskip
We thank Nicola Bennert for her assistance in the reduction of 
Q+A spectra, and Zhaohui Shang and Rajib Ganguly for providing the original
version of Fig.~\ref{qaimg}.
This work was supported in part under proposal GO-10421 by NASA
through a grant from the Space Telescope Science Institute, which is 
operated by the Association of Universities for Research in Astronomy, Inc.,
under contract NAS5-26555.   Additional support was provided by the 
National Science Foundation, under grant number AST 0507450.  
Some of the data presented herein were obtained at the W.M. Keck Observatory, 
which is operated as a scientific partnership among the California Institute 
of Technology, the University of California and the National Aeronautics and 
Space Administration. The Observatory was made possible by the generous 
financial support of the W.M. Keck Foundation.


\end{document}